\documentclass[onecolumn,tighten,times]{aastex631}

\usepackage{amsmath}
\usepackage{bm}

\shorttitle{Molecular Gas in ULIRGs}
\shortauthors{Farrah et al.}

\newcommand{\uhm}{Department of Physics and Astronomy, University of Hawai`i at M\=anoa, 2505 Correa Rd., Honolulu, HI, 96822, USA}

\begin{document}

\title{Molecular Gas Heating, Star Formation Rate Relations, and AGN Feedback in Infrared-luminous Galaxy Mergers}

\correspondingauthor{Duncan~Farrah}
\email{dfarrah@hawaii.edu}

\author[0000-0003-1748-2010]{Duncan~Farrah}
\affiliation{\uhm}
\affiliation{Institute for Astronomy, University of Hawai`i,  2680 Woodlawn Dr., Honolulu, HI, 96822, USA}

\author[0000-0002-2612-4840]{Andreas Efstathiou}
\affiliation{School of Sciences, European University Cyprus, Diogenes Street, Engomi, 1516 Nicosia, Cyprus}

\author[0000-0002-9149-2973]{Jose Afonso}
\affiliation{Instituto de Astrof\'{i}sica e Ci\^{e}ncias do Espa\c co, Universidade de Lisboa, Portugal}
\affiliation{Departamento de F\'{i}sica, Faculdade de Ci\^{e}ncias, Universidade de Lisboa, Portugal}

\author[0000-0002-9548-5033]{David L Clements}
\affiliation{Imperial College London, Blackett Laboratory, Prince Consort Road, London, SW7 2AZ, UK}

\author[0000-0002-6917-0214]{Kevin~S.~Croker}
\affiliation{\uhm}

\author[0000-0003-0917-9636]{Evanthia Hatziminaoglou}
\affiliation{ESO, Karl-Schwarzschild-Str 2, D-85748 Garching bei München, Germany}

\author{Maya Joyce}
\affiliation{Department of Physics and Astronomy, Michigan State University, East Lansing, MI 48824, USA}

\author[0000-0002-7716-6223]{Vianney Lebouteiller}
\affiliation{AIM, CEA, CNRS, Universite Paris-Saclay, Universite Paris Diderot, Sorbonne Paris Cite, F-91191 Gif-sur-Yvette, France}

\author{Aláine Lee}
\affiliation{Institute for Astronomy, University of Hawai`i,  2680 Woodlawn Dr., Honolulu, HI, 96822, USA}

\author{Carol Lonsdale}
\affiliation{National Radio Astronomy Observatory, Charlottesville, VA, USA}

\author[0000-0001-6139-649X]{Chris Pearson}
 \affiliation{RAL Space, STFC Rutherford Appleton Laboratory, Didcot, Oxfordshire OX11 0QX, UK}
 \affiliation{The Open University, Milton Keynes MK7 6AA, UK}
 \affiliation{Oxford Astrophysics, University of Oxford, Keble Rd, Oxford OX1 3RH, UK}

\author[0000-0003-0624-3276]{Sara~Petty}
\affiliation{NorthWest Research Associates, 3380 Mitchell Ln., Boulder, CO 80301, USA}
\affiliation{Convent \& Stuart Hall Schools of the Sacred Heart, 2222 Broadway, San Francisco, CA 94115, USA}

\author[0000-0002-5206-5880]{Lura K Pitchford}
\affiliation{Department of Physics and Astronomy, Texas A\&M University, College Station, TX, USA} 
\affiliation{George P. and Cynthia Woods Mitchell Institute for Fundamental Physics and Astronomy, Texas A\&M University, College Station, TX, USA}

\author[0000-0001-6854-7545]{Dimitra Rigopoulou}
\affiliation{Oxford Astrophysics, University of Oxford, Keble Rd, Oxford OX1 3RH, UK}

\author[0000-0002-0730-0781]{Aprajita Verma}
 \affiliation{Oxford Astrophysics, University of Oxford, Keble Rd, Oxford OX1 3RH, UK}

\author[0000-0002-6736-9158]{Lingyu Wang}
\affiliation{Kapteyn Astronomical Institute, University of Groningen, Postbus 800, 9700 AV Groningen, the Netherlands}
\affiliation{SRON Netherlands Institute for Space Research, Landleven 12, 9747 AD, Groningen, the Netherlands}

\begin{abstract}
We examine the origin of molecular gas heating in a sample of 42 infrared-luminous galaxies at $z<0.3$ by combining two sets of archival data. First, integrated CO line luminosities in the 1-0 and 5-4 through 13-12 transitions. Second, results from radiative transfer modelling that decompose their bolometric emission into starburst, AGN, and host galaxy components. We find that the CO 1-0 and 5-4 through 9-8 lines primarily arise via radiative heating in the starburst and the host galaxy. In contrast, the CO 10-9 through 13-12 lines may arise primarily in the starburst and AGN, with an increasing  contribution from mechanical heating and shocks. For the sample as a whole, we find no evidence that AGN luminosity affects the heating of molecular gas by star formation. However, for starbursts with low initial optical depths, a more luminous AGN may reduce the efficiency of starburst heating of the CO 5-4 and above lines, consistent with negative AGN feedback. 
\end{abstract}

\section{Introduction}
A fundamental channel for galaxy assembly at $z\gtrsim1$ is the conversion of molecular gas to stellar and supermassive black hole (SMBH) mass in brief ($\sim10^{8}$\,yr) but intense episodes of activity. These episodes, which are almost invariably heavily obscured and thus most luminous at infrared wavelengths, can harbor star formation rates (SFRs) of up to several thousand Solar masses per year \citep{mrr18}, and can increase SMBH masses by up to $\sim1$ dex \citep{farrah22}.  At low redshifts, infrared-luminous galaxies are rare, but are almost invariably observed to be mergers \citep{mur96,gen01,farrah01,bush02,haan11,kim13,petty14} and usually harbor both intense starbursts and luminous AGN \citep{gen98,farrah03,far05,arm06,nard09,bern09,far13,far16,psan21}. Some fraction of local infrared-luminous galaxies later go through an optical QSO phase \citep{tacc02,farrah07fe,far09}. Studies of infrared-luminous galaxies at low-redshift are invaluable for understanding the physics of obscured star formation and AGN, and as a zeropoint for studying the redshift evolution of this population. Infared-luminous galaxies at high redshift may have a lower merger fraction \citep{farrah02,kart12,far17,zavala18,gull19} but can also harbor higher SFRs and more luminous AGN  \citep{far02sub,alex05,hatz05,heinis13,buss15,harris16,pitch16,simp17,bourne17,marr18,mrr18,stach18,malek18,pitch19,wangl21,gao21,efs21}. Reviews of the infrared-luminous phases of galaxy assembly and extragalactic infrared astronomy can be found in \citep{sm96,blain02,lon06,cas14,farrah19,perez21,sajin22}.

The heating of molecular gas in infrared-luminous galaxies is thus an important diagnostic of both the conversion of molecular gas to stellar and SMBH mass, and of the impact of starburst and AGN activity on the interstellar medium (ISM) of their host galaxies \citep{solv05,cicone14,george15,tombesi15,ishibashi22}. While the bulk of the molecular gas in galaxies is in the form of 'cold' $H_{2}$, this molecule's lack of a dipole means that cold $H_{2}$ is challenging to observe directly\footnote{In contrast, 'warm' $H_{2}$ can be observed directly in the near- and mid-infrared \citep{higdon06,bern09}, but comprises only a small fraction of the total molecular gas mass.}. Instead, the molecular gas is usually traced via observations of rotational transitions of carbon monoxide ($^{12}CO$, CO hereafter), since CO has both a strong dipole moment and relatively low-lying energy transitions. The ground state transition (J=1-0) and transitions up to about J=4-3 of CO are good tracers of the total molecular gas reservoir in galaxies \citep{solom97,greve05,carilli13,bolatt13,gen15,sai17,yamashita17,molina19}. The higher J transitions trace the warmer, denser molecular gas associated with starbursts and AGN \citep[e.g.][]{weiss07,greve14,rosen15,mash15,pea16}. As a result, observations of CO emission in high-redshift galaxies have produced important insights into galaxy assembly processes \citep{scott11,aravena12,pitch19}.

Significant uncertainties remain over how different transitions of CO are produced. In principle, CO emission can arise in star-forming regions, AGN, or in the ISM heated by passively evolving stellar populations; One way to address this is to model physical conditions in the line-emitting gas using multiple transitions of CO simultaneously \citep{vandertak07,kamen18}. Another approach is to compare CO line luminosities to the total infrared luminosity of the galaxy \citep{greve14}. However, ambiguities in the power source behind the infrared emission persist in this approach.

In this paper, we take an alternative approach to determining the origin of CO heating in infrared-luminous galaxies. We take a sample of 42 gas-rich mergers with infrared luminosities in excess of $10^{12}$\,L$_{\odot}$ at $z<0.3$, for which we have completed radiative transfer modelling which constrains their starburst, AGN, and host galaxy luminosities. We then assemble archival observations of the 1-0 and 5-4 through 13-12 transitions of CO for this sample, and compare them to the component luminosities from the radiative transfer modelling to infer the origin of the CO gas heating as a function of rotational quantum number. Throughout, we assume \mbox{$H_0 = 70$\,km\,s$^{-1}$\,Mpc$^{-1}$}, \mbox{$\Omega = 1$}, and \mbox{$\Omega_{\Lambda} = 0.7$}. We convert all literature data to this cosmology where necessary.

\begin{table*}
\caption{The sample and their CO line luminosities. Units are $10^{6}$L$_{\odot}$. Uncertainties are $1\sigma$. All line luminosities are single-dish quantites. The origins of the data are given in \S\ref{sec:molgas}\label{tab:sample}}
\begin{center}
{\tiny
\begin{tabular}{lcccccccccc}
\hline
 \textbf{Name} & \textbf{1-0} & \textbf{5-4} & \textbf{6-5} & \textbf{7-6} & \textbf{8-7} & \textbf{9-8} & \textbf{10-9} & \textbf{11-10} & \textbf{12-11} & \textbf{13-12} \\
\hline
IRAS 00188-0856  &  $ 0.10\pm 0.01$  &  $ 12.9\pm 10.8$  &  $  9.3\pm  8.1$  &  $ 27.1\pm 10.1$  &  $ 35.4\pm  9.6$  &  $ 54.8\pm  9.6$  &  $ 39.1\pm  8.3$  &  $ 11.9\pm  9.2$  &  $ 41.1\pm  8.3$  &  ---         \\ 
IRAS 00397-1312  &  $ 2.12\pm 0.69$  &  $ 223.2\pm 44.1$  &  $ 138.3\pm 43.0$  &  ---          &  $ 105.3\pm 43.0$  &  $ 114.1\pm 43.5$  &  $ 147.1\pm 43.5$  &  ---          &  $ 181.9\pm 82.7$  &  $ 140.0\pm 44.1$ \\ 
IRAS 01003-2238  &  ---          &  ---          &  $ 15.2\pm  9.7$  &  $ 10.5\pm  5.6$  &  $ 10.4\pm  7.8$  &  $ 62.2\pm  7.7$  &  $ 42.1\pm 16.2$  &  $ 33.1\pm  7.1$  &  $ 22.8\pm  7.1$  &  $ 19.3\pm  7.1$ \\ 
IRAS 03158+4227  &  $ 0.40\pm 0.04$  &  $ 32.7\pm 11.8$  &  $ 58.7\pm 11.8$  &  $ 54.0\pm 12.0$  &  $ 56.2\pm 11.8$  &  ---          &  $ 54.2\pm 11.4$  &  $ 36.3\pm 11.4$  &  ---          &  ---         \\ 
IRAS 03521+0028  &  $ 0.48\pm 0.17$  &  $ 40.9\pm 29.4$  &  $ 67.2\pm 12.8$  &  $ 34.0\pm 16.3$  &  $ 59.7\pm 12.8$  &  ---          &  $ 72.9\pm 14.2$  &  $ 46.6\pm 12.8$  &  $ 36.1\pm 12.8$  &  $ 21.3\pm 14.2$ \\ 
IRAS 05189-2524  &  $ 0.13\pm 0.01$  &  $  5.2\pm  1.0$  &  $  7.1\pm  1.0$  &  $  8.8\pm  1.1$  &  $ 11.0\pm  1.0$  &  $ 11.0\pm  0.9$  &  $ 14.0\pm  1.4$  &  $ 10.1\pm  0.9$  &  $  8.4\pm  0.9$  &  $  8.1\pm  0.9$ \\ 
IRAS 06035-7102  &  $ 0.46\pm 0.09$  &  $ 41.5\pm  4.1$  &  $ 26.8\pm  4.1$  &  $ 32.7\pm  4.1$  &  $ 37.2\pm  4.1$  &  $ 41.1\pm  4.0$  &  $ 11.8\pm  3.9$  &  $ 34.2\pm  3.9$  &  $ 23.9\pm  3.9$  &  $ 10.0\pm  3.9$ \\ 
IRAS 06206-6315  &  $ 1.01\pm 0.20$  &  $ 16.8\pm 11.8$  &  $ 24.7\pm  5.3$  &  $ 29.5\pm  5.5$  &  $ 19.8\pm  5.3$  &  $ 19.3\pm  5.3$  &  $ 21.9\pm  4.6$  &  $ 21.6\pm  4.6$  &  $  6.8\pm  5.2$  &  ---         \\ 
IRAS 07598+6508  &  $ 0.66\pm 0.01$  &  $ 28.5\pm 20.8$  &  $ 51.2\pm 12.1$  &  $ 15.7\pm  9.8$  &  $ 12.1\pm  8.4$  &  ---          &  $ 37.4\pm 27.5$  &  $ 35.8\pm 22.8$  &  $ 42.2\pm 20.5$  &  $ 29.0\pm 23.6$ \\ 
IRAS 08311-2459  &  ---          &  $ 70.2\pm  6.7$  &  $ 76.9\pm  6.7$  &  $ 57.1\pm  6.9$  &  $ 53.8\pm  6.7$  &  $ 34.7\pm  6.7$  &  $ 50.5\pm  6.0$  &  $ 41.0\pm  6.2$  &  $ 52.6\pm  6.2$  &  $ 16.9\pm  6.2$ \\ 
IRAS 08572+3915  &  $ 0.08\pm 0.01$  &  $  6.6\pm  4.7$  &  $ 19.6\pm  7.7$  &  ---          &  $  6.2\pm  3.2$  &  $ 18.4\pm  2.5$  &  $ 10.9\pm  2.3$  &  $ 13.5\pm  2.3$  &  $  5.6\pm  3.7$  &  $  4.1\pm  3.1$ \\ 
IRAS 09022-3615  &  ---          &  $ 33.4\pm  2.3$  &  $ 41.4\pm  2.3$  &  $ 44.1\pm  2.4$  &  $ 39.7\pm  2.3$  &  $ 39.4\pm  2.3$  &  $ 28.5\pm  2.2$  &  $ 26.5\pm  2.2$  &  $ 19.9\pm  2.2$  &  $ 16.4\pm  2.2$ \\ 
IRAS 10378+1109  &  ---          &  $ 21.8\pm 13.5$  &  $ 57.0\pm  9.8$  &  $ 63.5\pm  9.8$  &  $ 54.7\pm  9.8$  &  $ 27.7\pm  9.8$  &  $ 39.0\pm  9.0$  &  $ 21.5\pm  9.0$  &  $ 11.0\pm  8.4$  &  $ 46.3\pm  9.0$ \\ 
IRAS 10565+2448  &  $ 0.32\pm 0.03$  &  $ 13.3\pm  1.6$  &  $ 20.8\pm  1.6$  &  $ 16.0\pm  1.6$  &  $ 20.9\pm  1.6$  &  $ 15.2\pm  1.4$  &  $ 13.5\pm  1.4$  &  $ 10.9\pm  1.4$  &  $  5.6\pm  1.4$  &  $  5.4\pm  1.4$ \\ 
IRAS 11095-0238  &  $ 0.38\pm 0.12$  &  $ 33.4\pm  6.2$  &  $ 10.9\pm  8.1$  &  $ 25.2\pm  6.2$  &  $ 35.6\pm  6.2$  &  $ 31.2\pm  6.3$  &  $ 37.0\pm  6.1$  &  $ 18.5\pm  6.1$  &  $ 21.2\pm 14.2$  &  $ 23.3\pm  6.1$ \\ 
IRAS 12071-0444  &  ---          &  $ 41.2\pm  8.6$  &  $ 50.5\pm  8.6$  &  $ 26.6\pm  8.7$  &  $ 43.0\pm  8.6$  &  ---          &  $ 41.8\pm  8.0$  &  $ 23.1\pm  8.0$  &  $ 29.0\pm  8.0$  &  ---         \\ 
IRAS 13120-5453  &  $ 0.33\pm 0.02$  &  $ 25.8\pm  1.0$  &  $ 29.1\pm  1.0$  &  $ 30.5\pm  1.0$  &  $ 29.0\pm  1.0$  &  $ 21.8\pm  1.0$  &  $ 22.6\pm  1.0$  &  $ 14.8\pm  1.0$  &  $ 10.2\pm  1.0$  &  $  8.0\pm  1.0$ \\ 
IRAS 13451+1232  &  $ 0.47\pm 0.04$  &  $ 37.3\pm  9.1$  &  $ 16.4\pm 10.4$  &  $ 24.3\pm  9.0$  &  $ 13.0\pm  9.7$  &  ---          &  $ 29.9\pm  8.3$  &  $ 19.0\pm 13.3$  &  $ 16.6\pm 12.4$  &  $ 11.2\pm  8.7$ \\ 
IRAS 14348-1447  &  $ 0.83\pm 0.08$  &  $ 30.6\pm  4.8$  &  $ 31.3\pm  0.5$  &  $ 34.9\pm  5.0$  &  $ 49.6\pm  4.8$  &  $ 28.0\pm  4.9$  &  $ 41.3\pm  0.5$  &  $ 32.0\pm  0.5$  &  $ 14.7\pm  0.5$  &  ---         \\ 
IRAS 14378-3651  &  $ 0.26\pm 0.05$  &  $ 26.2\pm  3.4$  &  $ 18.8\pm  3.4$  &  $ 28.7\pm  3.5$  &  $ 19.7\pm  3.4$  &  $ 19.9\pm  3.2$  &  $ 16.3\pm  3.2$  &  $ 18.6\pm  3.2$  &  $ 14.9\pm  3.2$  &  $ 16.8\pm  3.2$ \\ 
IRAS 15250+3609  &  $ 0.09\pm 0.03$  &  $  7.9\pm  2.2$  &  $ 12.4\pm  2.2$  &  $ 13.4\pm  2.2$  &  $ 12.6\pm  2.2$  &  $  8.9\pm  2.2$  &  $  8.9\pm  2.2$  &  $  9.4\pm  2.2$  &  ---          &  ---         \\ 
IRAS 15462-0450  &  $ 0.16\pm 0.01$  &  $ 15.8\pm  5.3$  &  $ 10.9\pm  6.9$  &  $ 22.9\pm  5.4$  &  $ 13.7\pm  5.3$  &  ---          &  $ 19.2\pm  4.9$  &  $  7.1\pm  5.5$  &  $ 16.7\pm  4.9$  &  $ 17.1\pm  9.3$ \\ 
IRAS 16090-0139  &  $ 0.67\pm 0.22$  &  $ 45.5\pm  1.1$  &  $ 36.2\pm 10.5$  &  $ 114.2\pm 110.4$  &  $ 120.3\pm 10.5$  &  $ 163.0\pm 10.5$  &  $ 110.0\pm 10.5$  &  $ 72.6\pm 10.4$  &  $ 52.4\pm 10.4$  &  $ 54.3\pm 10.4$ \\ 
IRAS 17208-0014  &  $ 0.66\pm 0.07$  &  $ 25.3\pm  1.6$  &  $ 34.1\pm  1.6$  &  $ 35.5\pm  1.7$  &  $ 36.1\pm  1.6$  &  $ 31.2\pm  1.8$  &  $ 32.2\pm  1.9$  &  $ 17.8\pm  1.7$  &  $ 11.2\pm  1.7$  &  $  9.8\pm  1.7$ \\ 
IRAS 19254-7245  &  $ 0.51\pm 0.10$  &  ---          &  $ 10.4\pm  2.5$  &  $ 18.5\pm  2.7$  &  $ 16.7\pm  2.5$  &  $ 19.1\pm  2.6$  &  $ 16.7\pm  2.4$  &  $ 11.4\pm  2.4$  &  $  7.3\pm  2.4$  &  $ 12.1\pm  2.4$ \\ 
IRAS 19297-0406  &  $ 0.51\pm 0.02$  &  $ 29.7\pm  5.4$  &  $ 27.3\pm  5.4$  &  $ 32.1\pm  5.7$  &  $ 43.9\pm  5.4$  &  $ 37.3\pm  5.5$  &  $ 13.2\pm  0.6$  &  $ 22.2\pm  0.6$  &  $  8.4\pm  8.4$  &  $ 20.6\pm  5.5$ \\ 
IRAS 20087-0308  &  $ 0.88\pm 0.03$  &  $ 65.5\pm  6.7$  &  $ 29.5\pm  6.7$  &  $ 56.0\pm  6.9$  &  $ 40.0\pm  6.7$  &  $ 27.6\pm  6.7$  &  $ 48.6\pm  6.6$  &  $ 31.2\pm  6.6$  &  $ 21.1\pm 12.7$  &  $ 30.6\pm  6.6$ \\ 
IRAS 20100-4156  &  $ 0.48\pm 0.04$  &  $ 68.5\pm 11.1$  &  $ 19.3\pm 15.4$  &  $ 47.2\pm 11.6$  &  $ 65.1\pm 11.1$  &  ---          &  $ 93.7\pm 11.4$  &  $ 37.2\pm 11.4$  &  $ 42.9\pm 25.9$  &  $ 28.7\pm 16.9$ \\ 
IRAS 20414-1651  &  $ 0.17\pm 0.06$  &  $ 12.2\pm  8.7$  &  $  7.2\pm  5.0$  &  $ 12.9\pm  4.8$  &  $ 19.4\pm  4.6$  &  $ 16.5\pm  4.6$  &  $ 19.1\pm  4.5$  &  $  6.3\pm  4.2$  &  $ 22.9\pm  4.4$  &  $ 10.2\pm  5.8$ \\ 
IRAS 20551-4250  &  $ 0.24\pm 0.05$  &  $ 17.5\pm  1.3$  &  $ 14.0\pm  1.3$  &  $ 20.0\pm  1.4$  &  $ 23.1\pm  1.3$  &  $ 16.5\pm  1.3$  &  $ 23.4\pm  1.3$  &  $ 21.3\pm  1.3$  &  $ 17.6\pm  1.3$  &  $ 15.6\pm  1.3$ \\ 
IRAS 22491-1808  &  $ 0.27\pm 0.03$  &  $ 35.5\pm  3.4$  &  $ 22.1\pm  3.4$  &  $ 21.1\pm  4.4$  &  $ 31.9\pm  3.4$  &  $ 21.6\pm  3.5$  &  $ 39.6\pm  3.7$  &  $ 31.4\pm  3.8$  &  $ 29.1\pm  3.8$  &  $ 26.3\pm  3.8$ \\ 
IRAS 23128-5919  &  $ 0.21\pm 0.04$  &  $ 12.7\pm  1.1$  &  $ 13.2\pm  1.1$  &  $ 19.1\pm  1.2$  &  $ 14.5\pm  1.1$  &  $ 14.3\pm  1.2$  &  $ 13.1\pm  1.2$  &  $ 10.6\pm  1.2$  &  $  8.2\pm  1.2$  &  $ 11.0\pm 11.0$ \\ 
IRAS 23230-6926  &  ---          &  $ 25.6\pm  6.3$  &  $  6.3\pm  6.3$  &  $ 30.1\pm  6.4$  &  $ 24.6\pm  6.3$  &  $ 35.7\pm  6.3$  &  $ 42.0\pm  5.9$  &  $ 23.9\pm  5.9$  &  $ 34.5\pm  5.9$  &  $ 27.6\pm  5.9$ \\ 
IRAS 23253-5415  &  ---          &  $ 17.4\pm 11.6$  &  $ 29.8\pm  8.5$  &  $ 30.5\pm  8.7$  &  $ 34.7\pm  8.5$  &  $ 43.3\pm  8.5$  &  $ 29.2\pm 23.2$  &  $ 17.3\pm  8.9$  &  $ 24.9\pm  8.9$  &  $ 23.9\pm  8.9$ \\ 
IRAS 23365+3604  &  $ 0.37\pm 0.06$  &  $  8.4\pm  2.9$  &  $ 22.1\pm  2.9$  &  $ 22.9\pm  3.0$  &  $ 14.5\pm  2.9$  &  $ 16.5\pm  2.9$  &  $ 26.4\pm  2.9$  &  $ 14.5\pm  2.9$  &  $ 17.6\pm  2.9$  &  $  8.9\pm  2.9$ \\ 
UGC 5101         &  $ 0.30\pm 0.03$  &  $ 11.0\pm  1.2$  &  $ 12.5\pm  1.2$  &  $  9.0\pm  1.2$  &  $ 10.4\pm  1.2$  &  $  7.7\pm  1.3$  &  $ 11.3\pm  1.3$  &  $  7.8\pm  1.3$  &  $  2.3\pm  1.3$  &  $  5.0\pm  1.3$ \\ 
Mrk 231          &  $ 0.35\pm 0.06$  &  $ 18.9\pm  1.3$  &  $ 18.8\pm  1.3$  &  $ 23.7\pm  1.4$  &  $ 28.0\pm  1.3$  &  $ 24.8\pm  1.3$  &  $ 28.8\pm  1.3$  &  $ 18.7\pm  1.3$  &  $ 16.8\pm  1.3$  &  $ 16.3\pm  1.3$ \\ 
Mrk 273          &  $ 0.25\pm 0.03$  &  $ 12.7\pm  0.7$  &  $ 16.2\pm  0.7$  &  $ 17.0\pm  0.7$  &  $ 15.8\pm  0.7$  &  $ 13.1\pm  0.8$  &  $ 13.8\pm  0.8$  &  $  7.6\pm  0.8$  &  $  5.1\pm  0.8$  &  $  5.5\pm  0.8$ \\ 
Mrk 463          &  $ 0.11\pm 0.04$  &  $  2.6\pm  2.4$  &  $  5.7\pm  1.2$  &  $  4.5\pm  1.3$  &  $  3.9\pm  1.2$  &  $  2.2\pm  1.9$  &  $  2.3\pm  1.1$  &  $  4.0\pm  3.7$  &  $  3.3\pm  3.0$  &  ---         \\ 
Arp 220          &  $ 0.32\pm 0.03$  &  $ 15.3\pm  0.8$  &  $ 19.0\pm  0.8$  &  $ 19.8\pm  0.8$  &  $ 19.4\pm  0.8$  &  $ 18.0\pm  1.4$  &  $ 14.0\pm  1.5$  &  $ 11.0\pm  1.4$  &  $  6.6\pm  1.4$  &  $  4.4\pm  1.4$ \\ 
NGC 6240     &  $ 0.43\pm 0.04$  &  $ 33.8\pm  1.0$  &  $ 38.2\pm  1.0$  &  $ 43.5\pm  1.0$  &  $ 43.2\pm  1.0$  &  $ 36.4\pm  1.0$  &  $ 33.6\pm  1.0$  &  $ 28.9\pm  1.0$  &  $ 22.7\pm  1.0$  &  $ 18.4\pm  1.0$ \\ 
Mrk 1014         &  $ 0.35\pm 0.03$  &  $ 63.6\pm 14.3$  &  $ 87.1\pm 14.3$  &  $ 37.5\pm 14.5$  &  ---          &  ---          &  $ 22.2\pm 17.6$  &  $ 16.6\pm 10.3$  &  ---          &  $ 30.4\pm 19.7$ \\ 
\hline
\end{tabular}
}
\end{center}
\end{table*}

\section{Methods}

\subsection{Sample Selection}
The sample selection is described in full in \citep{efs22,farrah22}, and is summarized here. The sample includes ultraluminous infrared galaxies (ULIRGs, systems with rest-frame $1-1000\mu$m luminosities in excess of $10^{12}$L$_{\odot}$) with IRAS 60$\mu$m fluxes greater than $\sim$2Jy that were assembled for the HERschel ULIRG Reference Survey \citep{far13,spoon13,pea16,cle18}.  This sample is representative of ULIRGs in the low-redshift Universe as it comprises nearly all known systems with $L_{IR}<10^{12}$L$_{\odot}$ at $z<0.27$. The sample spans approximately one order of magnitude in total infrared luminosity, includes merger stages from early to late stage, and all four primary optical spectral classes (HII, LINER, Sy2, Sy1).  The host galaxy properties are typical of massive, gas-rich galaxies in the local Universe. These data and diagnostic plots are given in  \citep{efs22,farrah22}.

\subsection{Molecular Gas Data}\label{sec:molgas}
We use the CO 1-0 transition, which traces cold molecular gas, and the CO 5-4 through 13-12 transitions, which trace hotter molecular gas.  We explored the possibility of using the 2-1 through 4-3 lines, and in lines above 13-12. There are, however, insufficient observations of our sample in these lines to enable a statistical analysis.  For CO 1-0 we use the compilation presented in \citet{farrah22}. These data are mostly taken from \citet{kamen16}, and/or a wider set of archival observations \citep{san89,mira90,downes93,solom97,baan08,chung09,grev09,braun11,papa12,xia12,meij13,ueda14,mash15,sliwa17,gowa18,ruff18,brown19,herr19,foto19,tan21}. For the 5-4 through 13-12 lines, we assemble data from \citet{pea16}, who present a homogeneous analysis of these lines as observed by the SPIRE instrument onboard {\itshape Herschel}. Data for a few objects are also taken from \citet{kamen16}. In all cases the CO line observations are spatially unresolved. Spatially resolved, interferometric observations of about a quarter of our sample exist, but we use spatially unresolved observations in all cases to give a consistent analysis.

\begin{figure*}
\centering
\includegraphics[width=12cm]{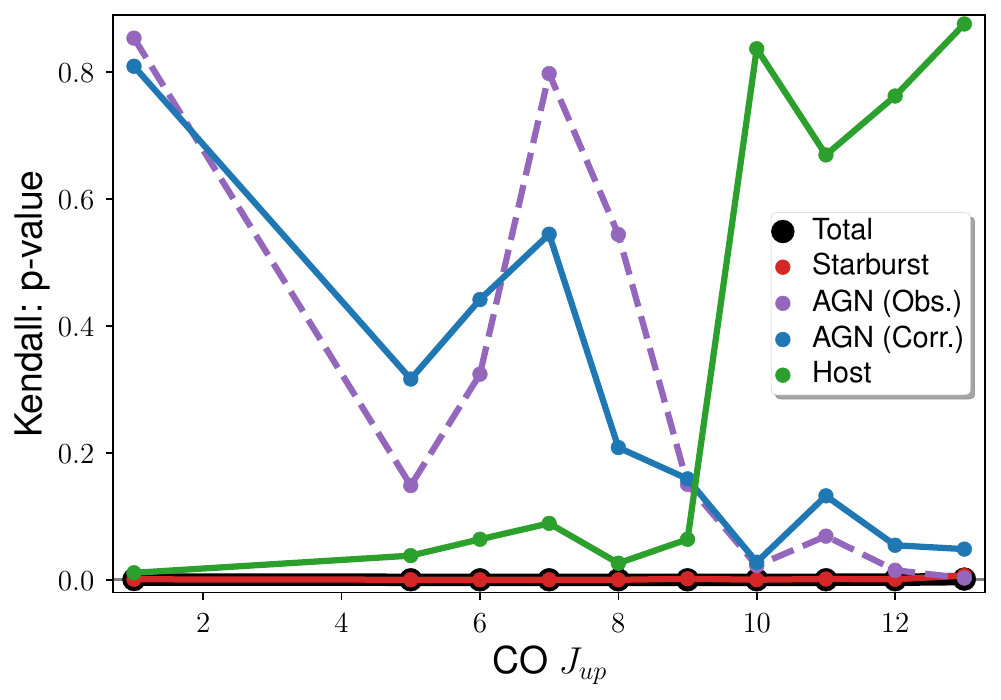} 
\caption{
The p-values from the Kendall-$\tau$ tests, comparing the CO luminosities to the total, starburst, AGN, and host infrared luminosities (\S\ref{sec:mgheat} \& \ref{sec:mgheatcomp}). A low p-value indicates that the hypothesis of no correlation can be rejected.\label{fig:pvals}}
\end{figure*}

\subsection{Infrared Emission Data}
We adopt the results from \citet{efs22}. This study decomposes the emission from our sample into contributions from a starburst, an AGN, and the host galaxy. The results from this study are consistent with previous SED modelling works, and provide derived physical parameters for each component. The complete set of parameters for each model, and their adopted ranges, are given in  Table 2 of  \citep{efs22}

The starburst model \citep{efstathiou09} assumes an exponentially decaying burst of star formation. The model includes a range of possible initial optical depths of the star-forming clouds, e-folding times of the starburst, and a range in starburst ages; thus tracking multiple possible evolutionary paths for the molecular clouds hosting the star formation. 

The AGN model  \citep{efsrr95,efstathiou95,efstathiou13} assumes a smooth, tapered disk in which the half-opening angle of the torus, the inclination angle of the torus relative to the observer, the ratio of inner to outer radius, and equatorial optical depth, can all vary. The AGN model also includes luminosities dervied by integrating the observed emission over $4\pi$ steradians, as well as luminosities that include a correction for the anisotropic structure of the obscurer. 

The host model assumes a S\'{e}rsic profile with $n=4$. The stars and dust are mixed, rather than the extinction applied as a foreground screen. The three parameters of the host model are the e-folding time of the star formation, the optical depth of the host, and the intensity of starlight. 

For our study, the key extracted physical quantities from the radiative transfer models are:

\begin{itemize}

\item $L^{o}_{tot}$: the total rest-frame infrared ($1-1000\mu$m) luminosity, uncorrected for anisotropic AGN emission.

\item $L_{Sb}$, $\dot{M}_{Sb}$: the rest-frame infrared luminosity of the starburst, and the SFR in the starburst averaged over the starbursts age, excluding the host. 

\item $L_{h}$, $\dot{M}_{h}$: the rest-frame infrared luminosity and SFR of the host galaxy, excluding the starburst.

\item $L^{o}_{AGN}$, $L^{c}_{AGN}$: Observed and anisotropy-corrected rest-frame infrared luminosity of the AGN. 

\item $\tau_{V}$: the initial optical depth of giant molecular clouds in the starburst, with an allowed range of $50 < \tau_{V} < 250$.

\end{itemize}

\noindent A description of these quantities is given in \citet{efs22}.

\begin{figure*}
\centering
\includegraphics[width=12cm]{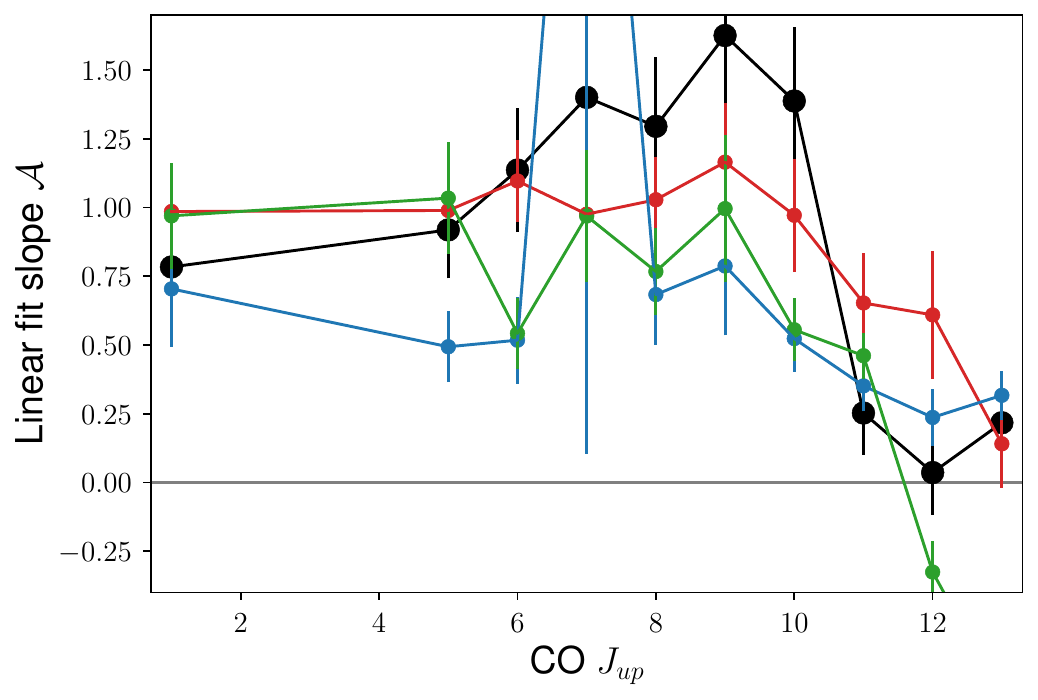} 
\caption{
The slopes of the fits between CO luminosity and total, starburst, host, and AGN luminosity, using Equation \ref{eq:ltoteq} (\S\ref{sec:mgheat} \& \ref{sec:mgheatcomp}). The colors follow the Figure \ref{fig:pvals} legend.\label{fig:allslopes3}}
\end{figure*}

\section{Results \& Discussion}\label{sec:res}

\subsection{Correlations with Total Infrared Luminosity}\label{sec:mgheat}
We first examine the relations between the CO luminosities and the total observed infrared luminosities ($L^{o}_{tot}$). Using the Kendall-$\tau$ test, we find significant correlations ($\tau > 0.38$, $p<0.01$) between all CO line luminosities and $L^{o}_{tot}$. The $p$-values are plotted as the black line in Figure \ref{fig:pvals}.

We next derive relations between CO luminosity and $L^{o}_{tot}$, of the form: 

\begin{linenomath}
\begin{equation}\label{eq:ltoteq}
\log(L_{CO}) = \mathcal{A}\log(L^{o}_{tot}) + \mathcal{B}
\end{equation}
\end{linenomath}

\noindent We used the Orthogonal Distance Regression algorithm \citep{bog87} as implemented within the {\itshape SciPy} Python library using a random seed of 2001. This yields:

\begin{eqnarray}
\log(L_{1-0})   & = & 0.78\pm0.17\log(L^{o}_{Tot}) - 4.02\pm2.07 \\
\log(L_{5-4})   & = & 0.92\pm0.18\log(L^{o}_{Tot}) - 3.78\pm2.17 \\
\log(L_{6-5})   & = & 1.14\pm0.23\log(L^{o}_{Tot}) - 6.44\pm2.76 \\
\log(L_{7-6})   & = & 1.40\pm0.23\log(L^{o}_{Tot}) - 9.65\pm2.79 \\
\log(L_{8-7})   & = & 1.30\pm0.25\log(L^{o}_{Tot}) - 8.33\pm3.08 \\
\log(L_{9-8})   & = & 1.63\pm0.33\log(L^{o}_{Tot}) - 12.38\pm3.99 \\
\log(L_{10-9})  & = & 1.39\pm0.27\log(L^{o}_{Tot}) - 9.50\pm3.30 \\
\log(L_{11-10}) & = & 0.25\pm0.15\log(L^{o}_{Tot}) + 4.31\pm1.86 \\
\log(L_{12-11}) & = & 0.04\pm0.15\log(L^{o}_{Tot}) + 6.79\pm1.88 \\
\log(L_{13-12}) & = & 0.22\pm0.17\log(L^{o}_{Tot}) + 4.56\pm2.01 
\end{eqnarray}

\begin{figure*}
\centering
\includegraphics[width=16cm]{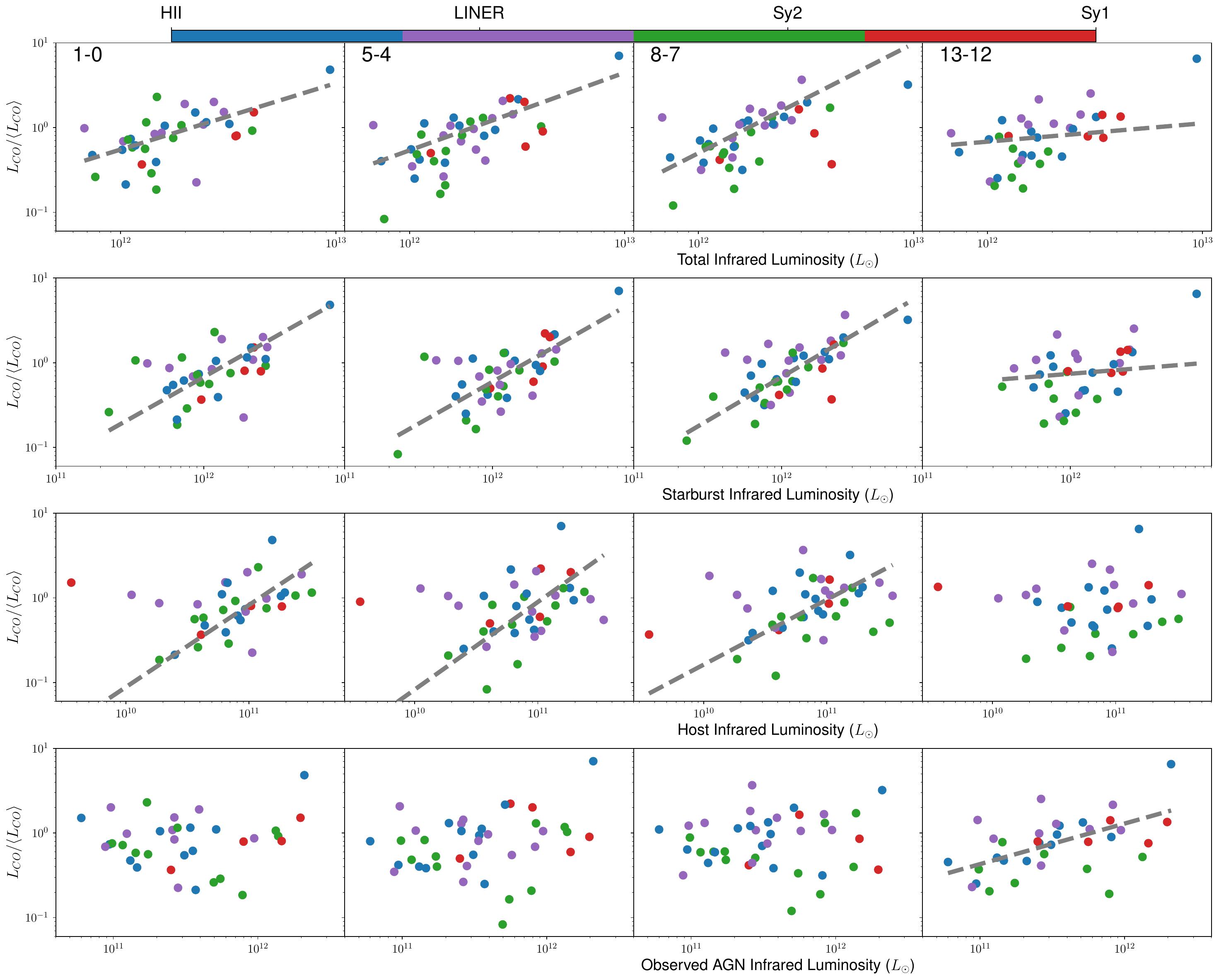} 
\caption{Example relations between CO line luminosities $L_{CO}$ (Table \ref{tab:sample}), normalized by the relevant mean CO luminosity $\langle L_{CO} \rangle$, and total, starburst, host, and observed AGN infrared luminosities (\S\ref{sec:mgheat} \& \ref{sec:mgheatcomp}). The best-fit relations are plotted, if there is evidence for a correlation. No strong dependence on optical spectral type is observed.\label{fig:manres1}}
\end{figure*}

\noindent (see also e.g. \citealt{greve14}). The slopes as a function of J$_{up}$ are plotted in Figure \ref{fig:allslopes3}. The slopes are consistent with a linear relation between $L_{CO}$ and $L^{o}_{tot}$ for J=10-9 and below, but a sub-linear, possibly flat relation for J=11-10 and above. Some example fits are presented in Figure \ref{fig:manres1}.

\subsection{Component Luminosity Correlations}\label{sec:mgheatcomp}
We next evaluate the Kendall-$\tau$ coefficients for each CO luminosity against $L_{Sb}$, $L^{o}_{AGN}$, $L^{c}_{AGN}$, and $L_{h}$ (Figure \ref{fig:allslopes3}). All of the CO luminosities correlate with $L_{Sb}$. For $L_{h}$, $L_{1-0}$ and $L_{5-4}$ through $L_{9-8}$ show correlations, though they may be weaker than those with $L_{Sb}$. However, $L_{10-9}$ and above do not show correlations with $L_{h}$. The starburst and host SFRs exhibit similar behavior to $L_{Sb}$ and $L_{h}$.  For $L^{o}_{AGN}$ and $L^{c}_{AGN}$ the $L_{10-9}$ and above luminosities show a correlation, but $L_{9-8}$ and below do not (see also e.g. \citealt{espo22}). The $L^{o}_{AGN}$ and $L^{c}_{AGN}$ luminosities show identical behaviour in these correlations, consistent with the inclination of the AGN obscurer relative to the line of sight not playing a dominant role in determining the observed CO line luminosities. 

We next fit relations of the form in Equation \ref{eq:ltoteq} to those CO luminosities that show a correlation, for each of $L_{Sb}$, $L^{o}_{AGN}$, $L_{h}$, and $\dot{M}_{Sb}$. The results for $L_{Sb}$ are:

\begin{eqnarray} 
\log(L_{1-0})  &=& 0.99\pm0.16\log(L_{Sb}) - 6.35\pm1.92  \\
\log(L_{5-4})  &=& 0.99\pm0.16\log(L_{Sb}) - 4.59\pm1.89  \\
\log(L_{6-5})  &=& 1.10\pm0.15\log(L_{Sb}) - 5.85\pm1.79  \\
\log(L_{7-6})  &=& 0.98\pm0.12\log(L_{Sb}) - 4.36\pm1.44  \\
\log(L_{8-7})  &=& 1.03\pm0.16\log(L_{Sb}) - 4.99\pm1.88  \\
\log(L_{9-8})  &=& 1.16\pm0.22\log(L_{Sb}) - 6.66\pm2.61  \\
\log(L_{10-9}) &=& 0.97\pm0.21\log(L_{Sb}) - 4.36\pm2.50  \\
\log(L_{11-10})&=& 0.65\pm0.18\log(L_{Sb}) - 0.58\pm2.19  \\
\log(L_{12-11})&=& 0.61\pm0.23\log(L_{Sb}) - 0.09\pm2.80  \\
\log(L_{13-12})&=& 0.14\pm0.16\log(L_{Sb}) + 5.91\pm1.91    
\end{eqnarray}

\noindent The relations for $\dot{M}_{Sb}$ are:

\begin{eqnarray}
\log(L_{1-0})  &=& 1.18\pm0.18\log(\dot{M}_{Sb}) + 2.45\pm0.49 \\
\log(L_{5-4})  &=& 0.91\pm0.14\log(\dot{M}_{Sb}) + 5.00\pm0.38 \\
\log(L_{6-5})  &=& 0.97\pm0.13\log(\dot{M}_{Sb}) + 4.86\pm0.35 \\
\log(L_{7-6})  &=& 1.16\pm0.20\log(\dot{M}_{Sb}) + 4.34\pm0.56 \\
\log(L_{8-7})  &=& 0.95\pm0.14\log(\dot{M}_{Sb}) + 4.91\pm0.37 \\
\log(L_{9-8})  &=& 1.29\pm0.21\log(\dot{M}_{Sb}) + 3.95\pm0.58 \\
\log(L_{10-9}) &=& 1.02\pm0.17\log(\dot{M}_{Sb}) + 4.70\pm0.47 \\
\log(L_{11-10})&=& 0.89\pm0.18\log(\dot{M}_{Sb}) + 4.94\pm0.49 \\
\log(L_{12-11})&=& 0.86\pm0.24\log(\dot{M}_{Sb}) + 5.00\pm0.63 \\
\log(L_{13-12})&=& 0.74\pm0.23\log(\dot{M}_{Sb}) + 5.24\pm0.62 
\end{eqnarray}

\noindent For comparison, see e.g. \citep{hunt20,sanchez22}. The results for $L_{h}$ are:

\begin{eqnarray} 
\log(L_{1-0})  &=& 0.97\pm0.19\log(L_{h}) - 5.11\pm2.13  \\
\log(L_{5-4})  &=& 1.03\pm0.20\log(L_{h}) - 3.92\pm2.22  \\
\log(L_{6-5})  &=& 0.54\pm0.13\log(L_{h}) + 1.40\pm1.45 \\
\log(L_{7-6})  &=& 0.97\pm0.24\log(L_{h}) - 3.24\pm2.65  \\
\log(L_{8-7})  &=& 0.77\pm0.16\log(L_{h}) - 0.94\pm1.73  \\
\log(L_{9-8})  &=& 1.00\pm0.27\log(L_{h}) - 3.49\pm2.94  \\
\end{eqnarray}

\noindent The results for $L^{o}_{AGN}$ are:

\begin{eqnarray} 
\log(L_{10-9}) &=& 0.78\pm0.17\log(L^{o}_{AGN}) - 1.50\pm2.00  \\
\log(L_{11-10})&=& 0.63\pm0.15\log(L^{o}_{AGN}) + 0.06\pm1.76  \\
\log(L_{12-11})&=& 0.77\pm0.19\log(L^{o}_{AGN}) - 1.63\pm2.21  \\
\log(L_{13-12})&=& 0.48\pm0.12\log(L^{O}_{AGN}) + 1.70\pm1.44    
\end{eqnarray}

\noindent We plot some of these relations in Figure \ref{fig:manres1}.

\begin{figure*}
\centering
\includegraphics[width=16cm]{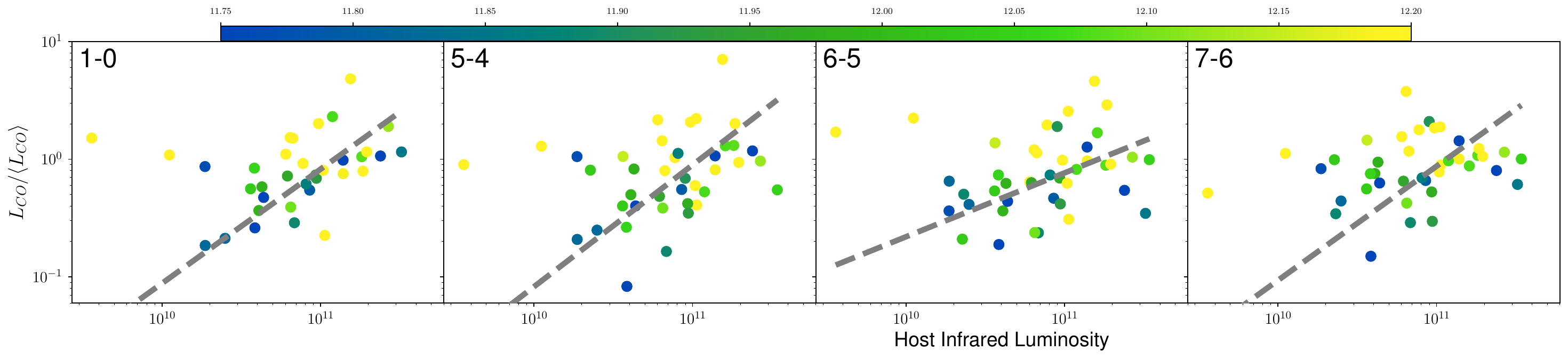} 
\caption{The $L_{CO} - L_{h}$ relations in Figure \ref{fig:manres1}, color-coded by $L_{Sb}$. The relations may tighten for lower starburst luminosities, consistent with the CO gas being heated by both host and starburst starlight.\label{fig:manres3}}
\end{figure*}

For $L_{1-0}$ and $L_{5-4}$ through $L_{9-8}$, the fits imply that the CO emission originates in the starburst and the host, with a smaller contribution from the AGN. Conversely, for $L_{10-9}$ and above, our results are consistent with the CO emission arising in the starburst and the AGN, with a smaller contribution from the host.  In qualitative consistency with this result is that the $L_{Sb} - L_{CO}$ relations may tighten for objects optically classified as HII regions, and the $L_{h} - L_{CO}$ relations may tighten for systems with lower starburst lumnosities (Figures \ref{fig:manres1} \&  \ref{fig:manres3}). 

For L$_{SB}$ and L$_{h}$ the slopes for the CO 9-8 and below relations are consistent with a linear relation.  For CO 10-9 and above, the relations with L$_{SB}$ are also consistent with linearity, though the slopes may start to flatten at CO 13-12.  In contrast, the slopes for the relations between CO 10-9 and above, and $L^{o}_{AGN}$, are all consistent with being sub-linear. 

We interpret these results as follows (see also e.g. \citet{greve14}). The CO 9-8 and below lines primarily arise from radiative heating in the starburst and the host, with a smaller contribution from the AGN. In the higher J transitions, host heating becomes insignificant. The emission in these CO lines primarily arises from starburst and AGN activity. In addition, mechanical heating and shocks become important for producing CO 11-10 and above.

\subsection{Multi-Component Fits}\label{sec:multifit}
To facilitate comparisons with galaxy evolution models, we present two-component fits of the form $L_{CO} = \alpha L_{A} + \beta L_{B} + \gamma$, where the luminosities $L_{A}$ and $L_{B}$ are chosen based on the results of the Kendall-$\tau$ tests; starburst and host of CO 9-8 and below, and starburst and AGN for CO 10-9 and above. This yields:

\begin{eqnarray}\label{eq:multi}
L_{1-0}   &=& (3.82\pm0.70)\frac{L_{Sb}}{10^{7}} + (4.93\pm1.58)\frac{L_{h}}{10^{6}} - (3.47\pm1.44)\times10^{5} \\
L_{5-4}   &=& (1.36\pm0.32)\frac{L_{Sb}}{10^{5}} + (1.57\pm0.40)\frac{L_{h}}{10^{4}} - (5.96\pm3.90)\times10^{6} \\
L_{6-5}   &=& (1.42\pm0.29)\frac{L_{Sb}}{10^{5}} + (8.57\pm2.02)\frac{L_{h}}{10^{5}} - (1.28\pm3.21)\times10^{6} \\
L_{7-6}   &=& (1.15\pm0.50)\frac{L_{Sb}}{10^{5}} + (5.30\pm1.79)\frac{L_{h}}{10^{4}} - (9.41\pm9.39)\times10^{6} \\
L_{8-7}   &=& (1.29\pm0.33)\frac{L_{Sb}}{10^{5}} + (1.78\pm0.41)\frac{L_{h}}{10^{4}} - (3.58\pm3.86)\times10^{6}\\
L_{9-8}   &=& (1.24\pm0.48)\frac{L_{Sb}}{10^{5}} + (1.27\pm0.50)\frac{L_{h}}{10^{4}} - (2.10\pm5.24)\times10^{6} \\
L_{10-9}  &=& (1.16\pm0.51)\frac{L_{Sb}}{10^{5}} + (9.91\pm2.69)\frac{L^{o}_{AGN}}{10^{5}} - (7.20\pm7.61)\times10^{6} \\
L_{11-10} &=& (2.53\pm3.71)\frac{L_{Sb}}{10^{6}} + (6.08\pm1.53)\frac{L^{o}_{AGN}}{10^{5}} + (3.16\pm5.08)\times10^{6} \\
L_{12-11} &=& (2.37\pm3.79)\frac{L_{Sb}}{10^{6}} + (7.66\pm1.91)\frac{L^{o}_{AGN}}{10^{5}} - (3.00\pm5.51)\times10^{6} \\
L_{13-12} &=& (-2.53\pm2.46)\frac{L_{Sb}}{10^{6}} + (2.12\pm0.70)\frac{L^{o}_{AGN}}{10^{5}} + (9.03\pm3.06)\times10^{6} 
\end{eqnarray}

\begin{figure*}
\centering
\includegraphics[width=8cm]{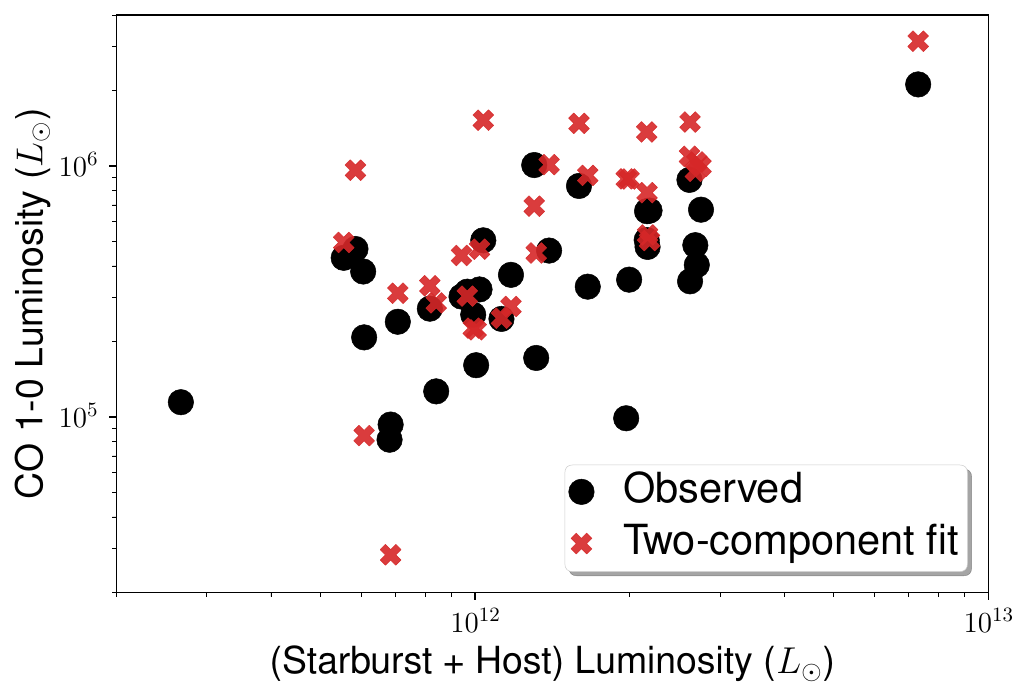} 
\includegraphics[width=8cm]{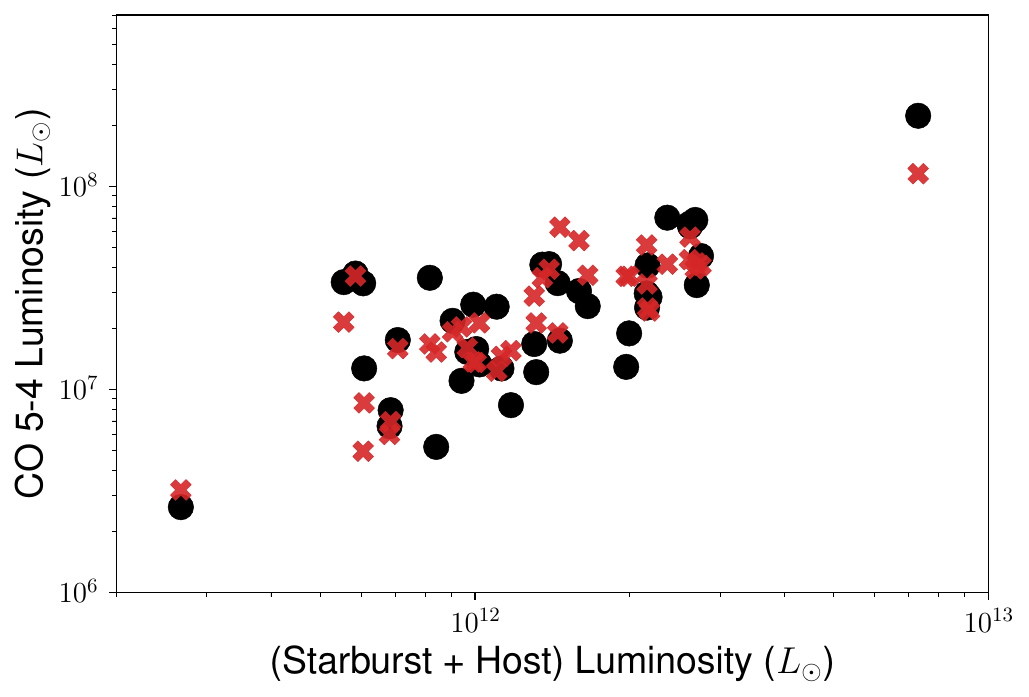}  \\
\includegraphics[width=8cm]{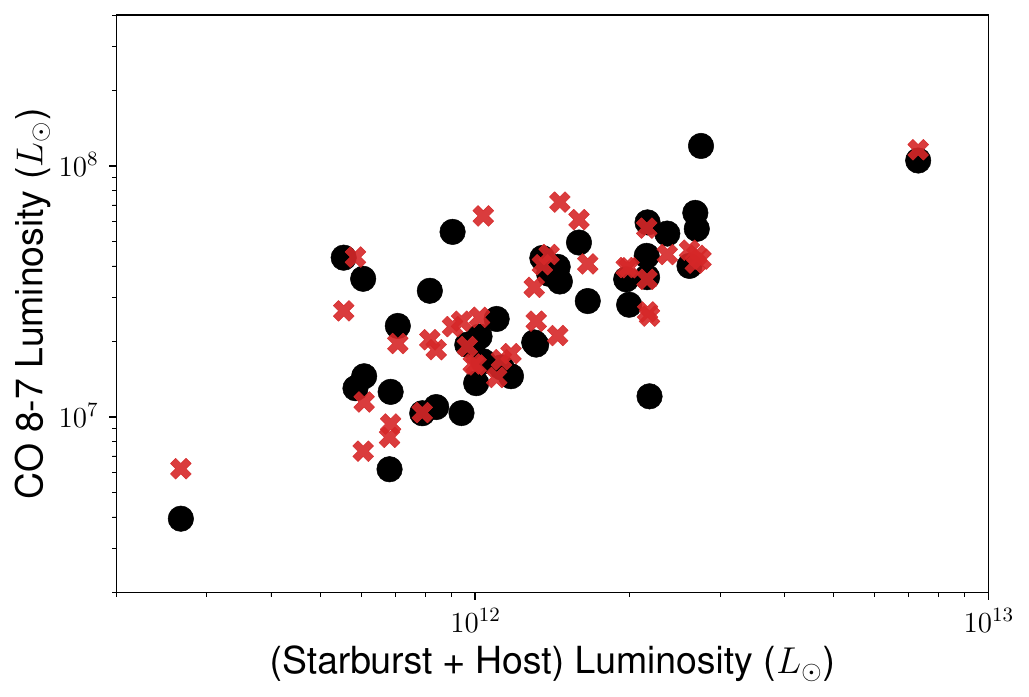} 
\includegraphics[width=8cm]{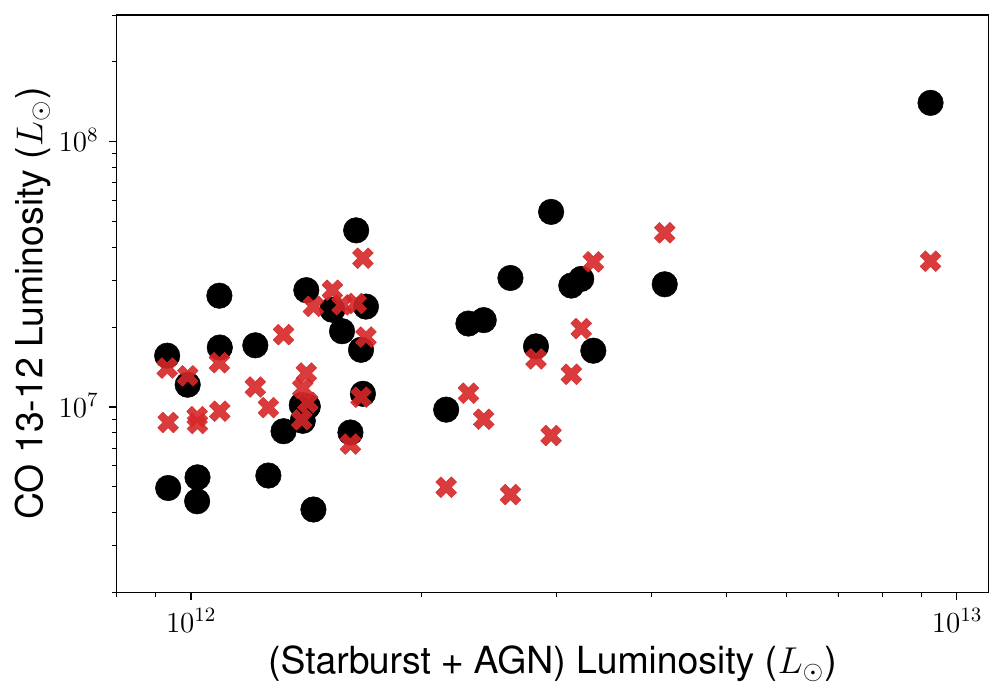} 
\caption{Example comparisons between CO luminosity and the sum of two component luminosites (starburst and host for 1-0, 5-4, and 8-7, starburst and AGN for 13-12). The red crosses show the results for a linear fit between the  two component luminosities and the CO luminosity, of the form $L_{CO} = \alpha L_{A} + \beta L_{B} + \gamma$ (\S\ref{sec:multifit}).\label{fig:multi}}
\end{figure*}

\noindent These fits are consistent with the log-linear fits, though they suggest the AGN may be the dominant contributor to the CO 11-10 and above emission. The fits are given in Figure \ref{fig:multi}. We do not present fits of the form: $L_{CO} = \alpha L_{Sb} + \beta L_{AGN} + \gamma L_{h}$ as they proved challenging to constrain with the number of objects in our sample.

\subsection{Star Formation Rate - Molecular Gas Scaling Relations}\label{sec:sfrh2}
As an alternative diagnostic of the relation between the cold ISM and star formation (see e.g. \citep{tacc20,sain22} for reviews) in our sample, we consider the relation between SFR and cold molecular hydrogen mass. To do so, we use the cold H$_{2}$ masses for our sample as summarized in \citet{farrah22}.   Since the total SFR is dominated in most cases by the starburst, we first fit a relation of the form in Equation \ref{eq:ltoteq}. We obtain:
 
\begin{linenomath}
\begin{equation}\label{eq:sfrh2single}
\log(M_{H_{2}}) = 1.11\pm0.19\log(\dot{M}_{Sb}) + 6.92\pm0.49 \\
\end{equation}
\end{linenomath}

\noindent This is consistent with the proposed universal scaling relation of \citet{lada12}. Instead using a two-component model yields:

\begin{linenomath}
\begin{equation}\label{eq:sfrh2multi}
M_{H_{2}} = (2.01 \pm 0.34)\times 10^{7} \dot{M}_{Sb} + (5.84 \pm 1.95)\times 10^{8} \dot{M}_{h} - (3.05 \pm 1.53)\times 10^{8} \\
\end{equation}
\end{linenomath}

\begin{figure*}
\centering
\includegraphics[width=12cm]{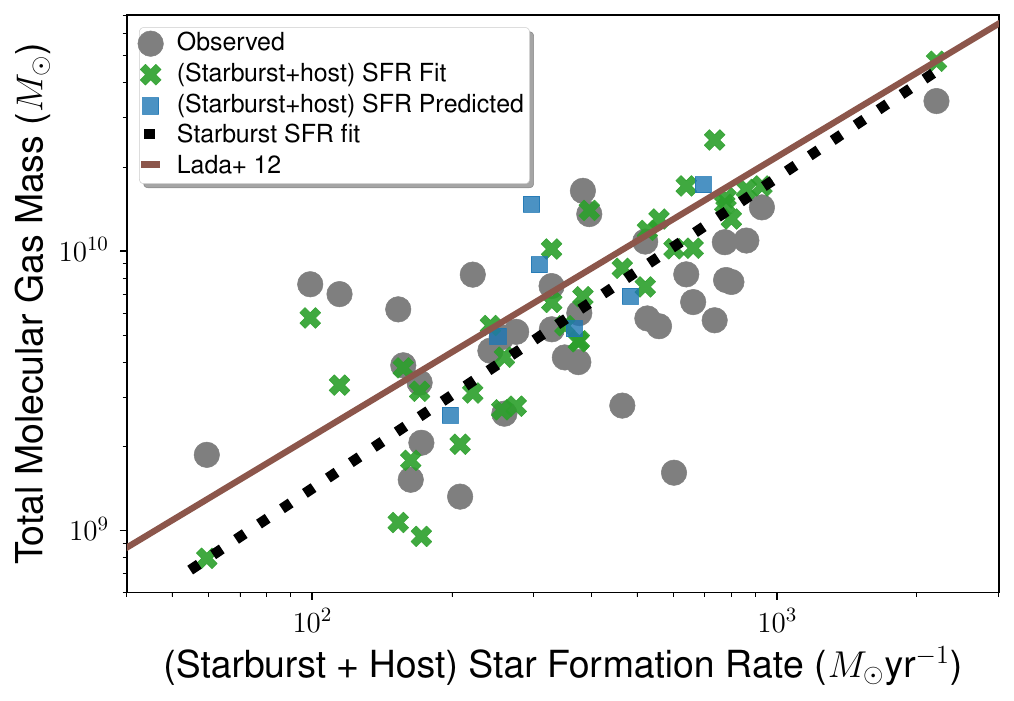} 
\caption{
The relation between total SFR and total molecular gas mass (\S\ref{sec:sfrh2}).  Overplotted are our starburst SFR - $M_{H_{2}}$ fit (Equation \ref{eq:sfrh2single}), our starburst+host SFR - $M_{H_{2}}$ fit (Equation \ref{eq:sfrh2multi}), and the predicted positions of objects without a CO 1-0 luminosity from Equation \ref{eq:sfrh2multi}.  Also plotted is the SFR-M$_{H_{2}}$ scaling relation from \citet{lada12}.\label{fig:h2sfr}}
\end{figure*}

\noindent We note though that Equation \ref{eq:sfrh2multi} does not give substantially reduced scatter relative to Equation \ref{eq:sfrh2single}. These results are summarized in Figure \ref{fig:h2sfr}.

\subsection{Starburst Efficiency}\label{sec:sbeff}
We next examine whether or not AGN luminosity can affect the heating of CO by the starburst. To do so, we consider the $L_{Sb}/L_{CO}$ ratio as a function of $L^{c}_{AGN}$, via the relation 

\begin{linenomath}
\begin{equation}\label{eq:effrel}
\log\left( \frac{L_{Sb}}{L_{CO}}\right) = \alpha\log{L^{c}_{AGN}} + \beta
\end{equation}
\end{linenomath}

\begin{figure*}
\centering
\includegraphics[width=8cm]{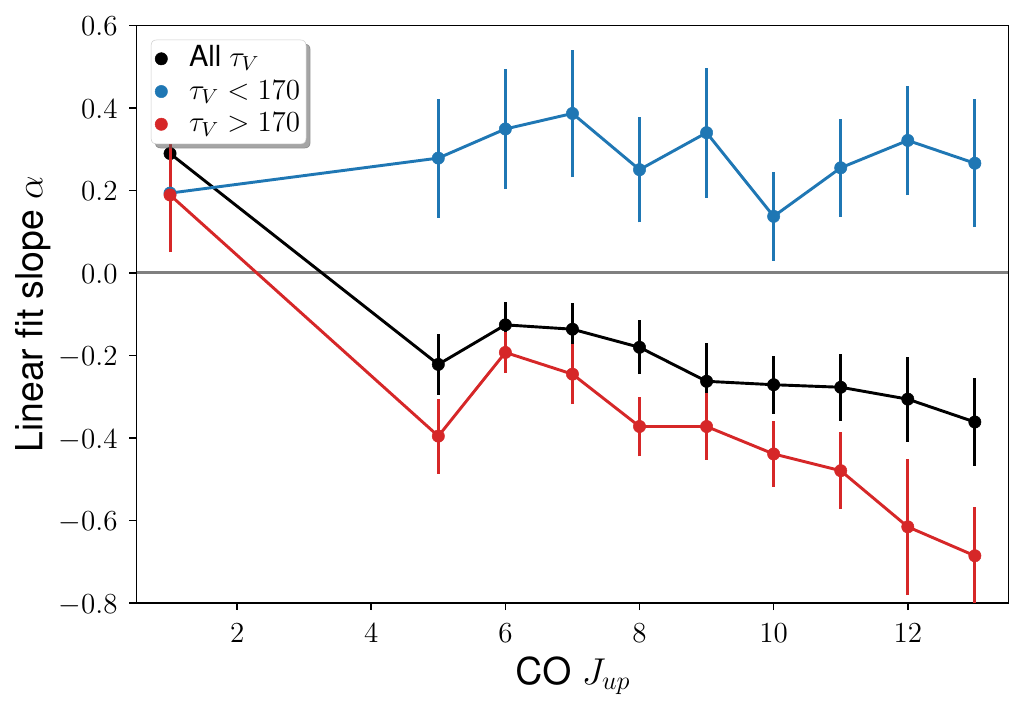}
\includegraphics[width=8cm]{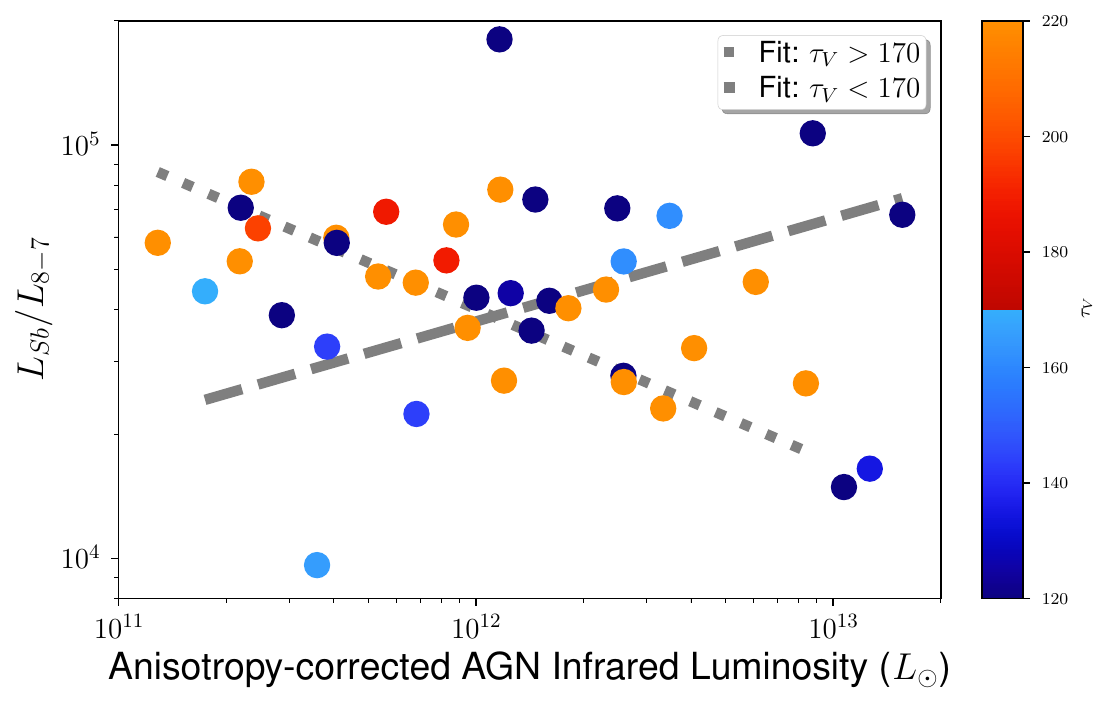}
\caption{{\itshape Left:} The slopes of the $L_{Sb}/L_{CO}$ vs. $L^{c}_{AGN}$ relations from Equation \ref{eq:effrel} (\S\ref{sec:sbeff}), plotted as a function of CO transition. The slopes are shown for the whole sample, and for the $\tau_{V}>170$ and $\tau_{V}<170$ sub-samples. {\itshape Right:}  An example of the $L_{Sb}/L_{CO}$ vs. $L^{c}_{AGN}$ relation, for $L_{8-7}$. Also shown are the individual fits for the $\tau_{V}>170$ and $\tau_{V}<170$ subsamples.}
\label{fig:coeffi2}
\end{figure*}

\noindent For the whole sample, we find a mildly negative relation between $L_{Sb}/L_{CO}$ and $L^{c}_{AGN}$ in most CO lines (Figure \ref{fig:coeffi2}). Combined with the results from \S\ref{sec:mgheat}, this can be interpreted as a significant contribution to heating the CO by the AGN in the CO 9-8 and above lines. We do not find that the AGN affects the starburst heating of the molecular gas, at least for the sample as a whole (see also e.g. \citealt{shang20agn,zhuang21,valerio21} but also \citealt{mcki21}). We note though that the factors that affect fueling of star formation in infrared-luminous mergers are subtle, and may vary by location within individual sources \citep[e.g.][]{thorp22}

Evidence for a relation between $L_{Sb}/L_{CO}$ and $L^{c}_{AGN}$ emerges when dividing the sample into two sub-samples at $\tau_{V}\sim170$ and fitting relations of the form in Equation \ref{eq:effrel}. The $\tau_{V} > 170$ sub-sample yields the same or slightly more negative slopes than the full sample - that is, similar trends in CO luminosity per unit starburst luminosity, as AGN luminosity increases. Conversely, the $\tau_{V} < 170$ sub-sample shows marginally positive slopes, suggesting less CO luminosity per unit starburst luminosity, as AGN luminosity increases. However, though the formal uncertainties imply this difference is highly significant, simulations that divide the sample in two equal sub-samples at random give an estimated confidence of $3\sigma$.

To interpret this result, we assume that the direct AGN contribution to producing CO emission is insignificant in all lines below CO 10-9, and sub-dominant for CO 10-9 and above. We also assume that the CO emission is not significantly absorbed. Our result is then consistent with more luminous AGN reducing the ability of star formation to produce warm and hot CO in starbursts with {\itshape lower} initial optical depths (see also e.g. \citealt{song21}). This effect can be interpreted as negative AGN feedback \citep{west12,fabian12,far12,bridge13,baron17,gonz17,gowa18,long18}, though the effect is subtle, and unlikely to be a viable channel to globally quench star formation in infrared-luminous galaxies. We speculate that the origin of the effect is that higher optical depth starbursts have higher ISM densities, which are harder to disrupt by a luminous AGN (see also e.g. \citep{mukh18,mack19}).

Other explanations seem less plausible. We do not see a dependence on the depth of the $9.8\mu$m silicate feature, or on the ratio of polycyclic aromatic hydrocarbon equivalent widths, suggesting that this effect does not arise in compact obsured nuclei \citep{garcia22}. More extincted starbursts could just be less luminous, but we see no such trend between $L_{Sb}$ and $\tau_{V}$. Alternatively, more extincted starbursts could be intrinsically more efficient at heating molecular gas. If this were true though then no trends with AGN luminosity would be expected.

\section{Conclusions}
We have studied the origin of CO emission in a sample of 42 infrared-luminous galaxy mergers at $z<0.27$ with infrared luminosities in excess of $10^{12}$\,L$_{\odot}$.  To do so, we take results from a recent radiative transfer modelling study of this sample, which decomposes their infrared luminosities into contributions from star formation, AGN activity, and the host galaxy. We then compare these component luminosities to archival measures of the integrated CO luminosities in the 1-0, and 5-4 through 13-12 transitions. Our conclusions are: 

\begin{enumerate}
	
\item The luminosities of the CO 1-0 and 5-4 through 13-12 lines are consistent with an origin in one or both of ongoing star formation, and heating by starlight in the host galaxy. We do not find evidence for a significant contribution to these lines from AGN activity. 

\item We present log-linear relations between the CO luminosities in the 1-0 and 5-4 through 13-12 transitions, and with the starburst and host galaxy luminosities. The slopes of these relations suggest an origin of the CO emission in gas heating. 

\item The luminosities of the CO 10-9 through 13-12 lines are consistent with an origin in one or both of the starburst and the AGN. We do not find evidence for a significant contribution from the host galaxy.  The slopes of the relations between these CO line luminosities and the starburst and AGN luminosities may be sub-linear, which is consistent with a contribution to these lines from mechanical heating and shocks. 
 
\item We also present two-component linear model fits between each CO line luminosity and infrared component luminosities. For CO 9-8 and below we fit against starburst and host luminosity. For CO 10-9 and above we fit against starburst and AGN luminosity. These fits are consistent with the log-linear model fits, and give straightforward conversions that can be used in galaxy simulations. 
 
\item For the sample as a whole we find no evidence for a relation between AGN luminosity and the $L_{CO}/L_{Sb}$, in any CO line. This suggests that a more luminous AGN does not reduce the efficiency by which the starburst heats the molecular gas. There is, however, evidence for a dependence on starburst initial optical depth ($\tau_{V}$).  Starbursts with low $\tau_{V}$ may heat the CO 5-4 transitions and above less efficiently. This is consistent with mild negative AGN feedback. 

\end{enumerate}

\begin{acknowledgments}
JA acknowledges financial support from the Science and Technology Foundation (FCT, Portugal) through research grants PTDC/FIS-AST/29245/2017, UIDB/04434/2020 and UIDP/04434/2020. 
\end{acknowledgments}

\bibliography{msbib}{}
\bibliographystyle{aasjournal}

\end{document}